\documentclass[reprint,amsmath,amssymb,aps,superscriptaddress, twocolumn]{revtex4-2}
\usepackage{graphicx}
\usepackage{bm}
\usepackage{float}
\usepackage{color}
\usepackage[dvipsnames]{xcolor}
\usepackage{hyperref}
\usepackage{graphicx}
\usepackage{braket}
\usepackage{appendix}
\usepackage{chemmacros}
\usepackage{natbib} 
\usepackage{siunitx}
\DeclareSIUnit\angstrom{\text {Å}}
\usepackage{amsmath}
\hypersetup{colorlinks=true, citecolor=blue, urlcolor=blue, linkcolor=blue}

\begin{document}

\preprint{APS/123-QED}

\title{Lower bound of the expressibility of ansatzes for Variational Quantum Algorithms}

\author{Tamojit Ghosh}\altaffiliation[tamojit2000@gmail.com]{}\affiliation{Department of Physics, Indian Institute of Technology Madras, Chennai-600036, Tamil Nadu, India}\affiliation{Department of Physical Sciences, Indian Institute of Science Education and Research Kolkata, Mohanpur- 741246, West Bengal, India}
 
 \author{Arijit Mandal}\altaffiliation[arijitmandal1997@gmail.com ]{}\affiliation{Department of Physics, Indian Institute of Technology Madras, Chennai-600036, Tamil Nadu,  India}\affiliation{Department of Physical Sciences, Indian Institute of Science Education and Research Kolkata, Mohanpur- 741246, West Bengal, India}

 \author{Shreya Banerjee}\altaffiliation[shreya93ban@gmail.com]{}
 \affiliation{Department of Physical Sciences, Indian Institute of Science Education and Research Kolkata, Mohanpur- 741246, West Bengal, India}
\affiliation{ Center for Quantum Science and Technology,
Siksha 'O' Anusandhan
University Bhubaneswar- 751030, Odisha, India}
\affiliation{Departement de physique and Institut quantique, Universit\'e de Sherbrooke, Sherbrooke- J1K 2R1, Qu\'ebec, Canada}

\author{Neetik Mukherjee}\altaffiliation[pchem.neetik@gmail.com]{}
 \affiliation{Department of Chemical Sciences, Indian Institute of Science Education and Research Kolkata, Mohanpur- 741246, West Bengal, India}
\affiliation{ Department of Chemistry,
Medi-Caps University, Indore- 453331, Madhya Pradesh, India}

\author{Prasanta K. Panigrahi}\altaffiliation[pprasanta@iiserkol.ac.in]{}
\affiliation{Department of Physical Sciences, Indian Institute of Science Education and Research Kolkata, Mohanpur- 741246, West Bengal, India}
\affiliation{ Center for Quantum Science and Technology,
Siksha 'O' Anusandhan
University Bhubaneswar- 751030, Odisha, India}

\begin{abstract}
The expressibility of an ansatz used in a variational quantum algorithm is defined as the uniformity with which it can explore the space of unitary matrices, i.e., its covering number. The expressibility of a particular ansatz has a well-defined upper bound \cite{originalpaper}. In this work, we show that the expressibility also has a well-defined lower bound in the hypothesis space. We provide an analytical expression for the lower bound of the covering number, which is directly related to expressibility. Further, we provide numerical analysis to support our claim. By calculating the bond length of hydrogen molecule ($H_2$) using different ansatzes in a variational quantum eigensolver (VQE) setting, we study the variation of equilibrium energy error with circuit depths. We show that in each ansatz template, a plateau exists for a range of circuit depths, which we call the \textit{set of acceptable points}, and the corresponding expressibility as the \textit{best expressive region}. We report that the width of this best expressive region in the hypothesis space is inversely proportional to the average error. Our analysis reveals that alongside trainability, the lower bound of expressibility also plays a crucial role in selecting variational quantum ansatzes.

\end{abstract}
\maketitle
\section{\label{sec:introduction}Introduction}
The promise of quantum computation to outperform classical computers in various tasks \cite{Montanaro2016,Harrow2017}, and effectively displaying `quantum advantage' \cite{preskillnisq} revolves around noiseless error-corrected quantum devices. However, algorithms that can be performed efficiently on today's Noisy Intermediate-Scale Quantum (NISQ) devices \cite{preskillnisq,Brooksnisq}, are also of significant interest.  Variational Quantum Algorithms (VQA) have turned out to be one of the best applications for near term quantum devices \cite{vqaadvantage1,vqeadvantage2,vqeadvantage3,nisq}, despite their limitations. They use parameterised quantum circuits in order to construct a quantum state subject to certain optimization routines which are performed on a classical computer. This act of deputising part of the routine to a classical computer helps keep the quantum circuits shallow and makes implementation viable on a NISQ device \cite{cerezo2021variational}.
One of the very first VQAs developed was the Variational Quantum Eigensolver (VQE) \cite{vqeadvantage2} which is designed to find the ground-state energy and the corresponding wavefunction of the Hamiltonian of a system. The variational principle in quantum mechanics says that given a Hamiltonian $H$ of a system and a trial wavefunction $\ket{\psi}$, the ground state energy $E_0$ of the system is upper bounded by the following relation:
\begin{equation*}
    E_0\leq\frac{\bra{\psi}H_0\ket{\psi}}{\braket{\psi|\psi}},
\end{equation*}
where equality is achieved if and only if the wavefunction $\ket{\psi}$ is equal to the ground state wavefunction. VQE uses the variational principle in Quantum Mechanics and starts with an initial parameterized trial wavefunction (the ansatz) \cite{vqereview}. It then finds the expectation value with respect to the Hamiltonian and optimizes the parameters in the wavefunction to minimize the energy at each step. The entire process is thus an optimization problem which (ideally) helps us obtain a good approximation to the ground state energy and the corresponding wavefunction. It has wide varying applications in Quantum Chemistry \cite{quantchem1}, condensed matter physics and material science \cite{condensed},  medical drug discovery \cite{vqedrug}, and nuclear physics \cite{vqenuclear}.

A crucial step in designing a Variational Quantum Algorithm is the selection of its initial ansatz. A wide variety of ansatzes can be found in the literature, such as hardware-efficient ansatzes \cite{hardwareefficient}, symmetry-preserving ansatzes \cite{symmansatze}, as well as problem-inspired ansatzes, e.g., Unitary Coupled Cluster (UCC) \cite{unitarycoupledansatz}, quantum alternating operator ansatz \cite{alternatingansatzqaoa}, variational Hamiltonian ansatz \cite{hamiltonianansatz}, variable structure ansatz \cite{adapt}. The Hardware-Efficient ansatzes (HEAs) are very versatile and can be easily tailored to the quantum device they are being used in. This feature and the ease of their construction led to their wide usage for various tasks \cite{practicaluseofhea,hardwareeffi1,hardwarefficient2}, whereas more general ansatzes, such as the quantum alternating operator ansatz \cite{alternatingansatzqaoa}, have also been proven to work well for certain problems. This leads to a practical need to determine the quality of each ansatz. One of such measures is the expressibility of an ansatz. Expressibility can be defined as the ability of an ansatz to uniformly explore the space of all possible unitaries ($\mathcal{U}(d)$). Formally, the expressibility of an ansatz is measured by the distance between the uniform distribution of unitaries sampled by changing the parameters in the ansatz and the uniform distribution of unitaries sampled from the maximally expressive Haar distribution\cite{zoeholmes,entangling,alternatingexp}. However, the expression for quantifying expressibility is provided by using the unitary t-design \cite{randomquantum}, which is expensive to compute and has several constraints. Other measures have also been developed to quantify expressibility that deal with particular ansatzes \cite{Abbas2021,huggins2019towards}.
The first step towards deriving an analytical expression for the expressibility of an ansatz in terms of its Hilbert space dimensions, the number of trainable gates, and the types of gates used was taken in \cite{originalpaper}. It uses the covering number\cite{coveringnumber} of the hypothesis space of the ansatz as a quantifier to obtain an upper bound for the expressibility of the said ansatz. The expressibility of an ansatz has an interesting relation with its performance. It was shown that, ansatzes like HEAs with a large circuit depth suffer from barren plateaus \cite{barren} due to their high expressibility \cite{zoeholmes}. It must be noted that vanishing gradients have been seen for shallow circuits too \cite{shallow}. This leads to trainability issues for these ansatzes \cite{barren,shallow,zoeholmes,dqnn,train1,rosa}

In this paper, we extend the results from \cite{originalpaper} to provide an expression for the lower bound for the expressibility of an ansatz, in terms of the circuit parameters. The physical implications of the lower bound of expressibility is significant in terms of ansatz designs. The existence of a lower bound indicates towards the lowest possible design of a circuit to achieve a good performance analytically, expressed in terms of qub(d)it numbers and number of trainable gates.   

This is an important result that, as far the best of our knowledge, has not been found yet. This helps us tighten the bound given by \cite{originalpaper} and brings us closer to a generic expression for expressibility. We take VQEs with Hardware Efficient Ansatz as a case study and perform numerical simulations of the ground state energy of a Hydrogen molecule using QISKIT. Different ansatz templates are compared according to their performance. The error of the VQE algorithm with respect to the  classical result (obtained by diagonalizing the Hamiltonian) is used as a quantifier of the performance of repeating layers of the ansatz templates. We find that each ansatz templates performs well for certain intermediate number of layers. We draw a relationship between the `best expressive region' of an ansatz template and the `average error' across this region. 

The presence of a lower bound to the expressibility of an ansatz used in VQE tells us that if our solution corresponds to a wavefunction in a low expressive region then an initial choice of a highly expressive trial ansatz will never be able to reach the solution leading to an erroneous VQE result. This shows, the under-performance of a highly expressive ansatz depends not only on the related  trainability issues \cite{zoeholmes}, but also with its intrinsic expressibility.

\section{\label{sec:preliminaries}Preliminaries}
\subsection{Variational Quantum Eigensolver}
\begin{figure*}[!hbt]
    \centering
   \includegraphics[scale=0.45, trim=0 3.5cm 0 2mm, clip]{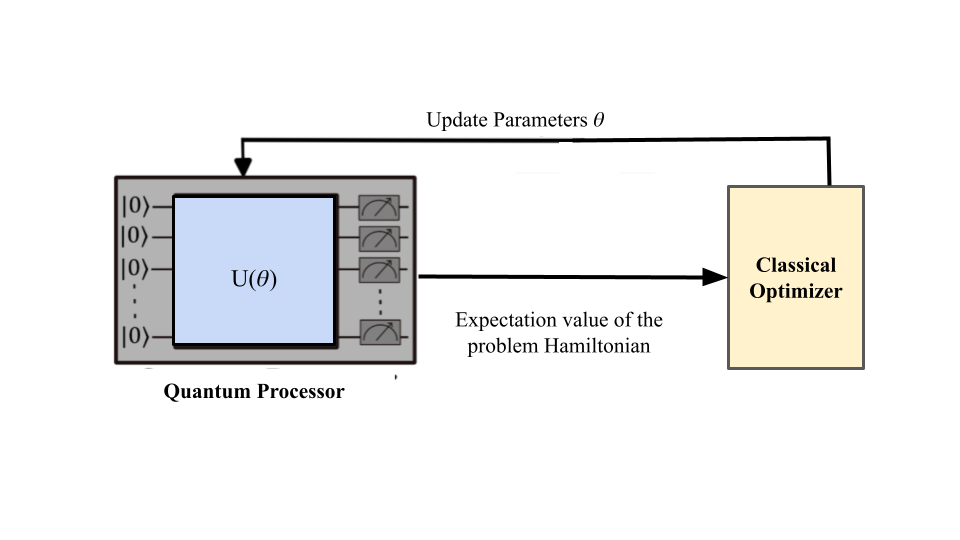}
    \caption{General workflow of Variational Quantum Algorithms (VQAs). The $\theta$ is a vector that contains all parameter values associated with the ansatz $U(\theta)$. The Expectation value of the problem Hamiltonian acts as the input to the classical optimizer. The output of the optimizer is a new $\theta$ vector that acts as an input to the quantum processor.}
    \label{fig:vqediagram}
\end{figure*}
 We consider a quantum state $\ket{\psi(\theta)}$, containing a set of tunable parameters $\theta$. $\ket{\psi(\theta)}$ can be obtained by applying a unitary $U(\theta)$ on a $N$ qubit register initialized to $\ket{\psi_0} = \ket{0}^{\otimes N}$. In a variational quantum eigensolver, the expectation value of the Hamiltonian $\hat{H}$ with respect to $\ket{\psi(\theta)}$ is calculated and minimized by optimizing the parameters $\theta$, with an aim of finding the ground state of $\hat{H}$. The VQE cost function can then be written as,

\begin{eqnarray}
    E_0^{VQE} &=& \min_{\theta}\bra{\psi_0}U^{\dagger}(\theta)\hat{H}U(\theta)\ket{\psi_0} \nonumber \\
    &=& \min_{\theta}\bra{\psi(\theta)}\hat{H}\ket{\psi(\theta)}.
\end{eqnarray}
As depicted in Fig.~\ref{fig:vqediagram}, the optimization of the expectation value of the problem Hamiltonian is classical. This hybrid algorithm tries to optimize (minimize) the expectation value of $\hat{H}$ with respect to a quantum state computed using quantum computer, by classically varying the parameters $\theta$. This is done by writing the Hamiltonian $\hat{H}$ as a weighted sum of spin operators.
\begin{equation}
    \hat{H}=\sum_{i=1}^k c_i \hat{P}_i
\end{equation}
where $c_i$ are the weights, $\hat{P}_i$ is a Pauli string on $N$ qubits i.e.,$\hat{P}_i\in\left\{I,X,Y,Z\right\}^{\otimes N}$ and $k$ denotes the number of such Pauli strings needed to represent the Hamiltonian.
\subsection{Covering number and the hypothesis space}
$Q$ is `$\epsilon$-covering' for a metric space $\mathcal{M}$ with distance $d$ if for every $y\in\mathcal{M}$ and $x\in Q$, $d(x,y)\leq\epsilon$. Then the cardinality $\mathcal{N}(\mathcal{M},d,\epsilon)$ of the smallest $Q$ is called ``$\epsilon$ covering number" of set $\mathcal{M}$ \cite{oldprl}.
\begin{equation}
    |Q|_{smallest}=\mathcal{N}(\mathcal{M},\epsilon,d)
\end{equation}
\par We now define the hypothesis space $\mathcal{H}$ as:
\begin{equation}
\label{eq:hypothesis}
    \mathcal{H}=\left\{Tr(\hat{U}(\theta)^\dag \hat{O}\hat{U}(\theta)\rho)|\theta\in \Theta\right\}
\end{equation}
where $\rho\in\mathbf{C}^{d^N\times d^N}$ is the $N$-qudit input quantum 
\vspace{0.05in}
state, $\hat{O}\in\mathbf{C}^{d^N\times d^N}$ is the problem Hamiltonian
\vspace{0.05in}
and $\hat{U}(\theta)=\prod_{i=1}^{N_g}\hat{u}_i(\theta_i) \in \mathcal{U}(d^N) $ is the 
\vspace{0.05in}
ansatz that we are using, with $N_g$ quantum gates. $\Theta$ is the parameter space that 
\vspace{0.05in}
contains the parameters $\theta_i$. $\mathcal{U}(d^k)$ denote the set of unitary matrices of dimension $d^k$, which also represents the set of all possible quantum gates used in the variational algorithm. 
\vspace{0.05in}
$\hat{u}_p(\theta_p)$ is the $p^{th}$ quantum gate with 
\vspace{0.05in}
$\hat{u}_p(\theta_p)\in\mathcal{U}\left(d^k\right)$ i.e.,it can 
\vspace{0.05in}
act on a maximum of $k$ qudits such that $k\leq N$. We choose the ansatz $\hat{U}(\theta)$ such that it has $N_g$ gates among which $N_{gt}$ are trainable (or parameterized) gates. 

Next, we define the operator group $\mathcal{H}_{circ}$ which consists of the different parameterised forms of a particular ansatz that we use in our variational algorithm.
\begin{equation}
    \mathcal{H}_{circ}=\{\hat{U}(\theta)^\dag \hat{O}\hat{U}(\theta)|\theta\in \Theta \}
\end{equation}

For $0<\epsilon\leq 1/10$, the $\epsilon$-covering number for the unitary group $\mathcal{U}(d^k)$ with respect to the operator norm distance $||\cdot||$, is bounded by the relation \cite{oldprl} :
\begin{equation}
\label{eq:Uupperlower}
    \left(\frac{3}{4\epsilon}\right)^{d^{2k}}\leq\mathcal{N}(\mathcal{U}(d^k),\epsilon,||\cdot||)\leq\left(\frac{7}{\epsilon}\right)^{d^{2k}}
\end{equation}
where the operator norm of an $n\times n$ matrix $A$ is defined as \cite{originalpaper}:
\begin{equation}
    \label{eq:opernorm}
    ||A|| =\sqrt{\mu_1\left(AA^\dag\right)}
\end{equation}
where $\mu_1\left(AA^\dag\right)$ is the largest eigenvalue of the matrix $AA^\dag$.

\begin{figure*}[hbt!]
    \centering
    \includegraphics[scale=0.5]{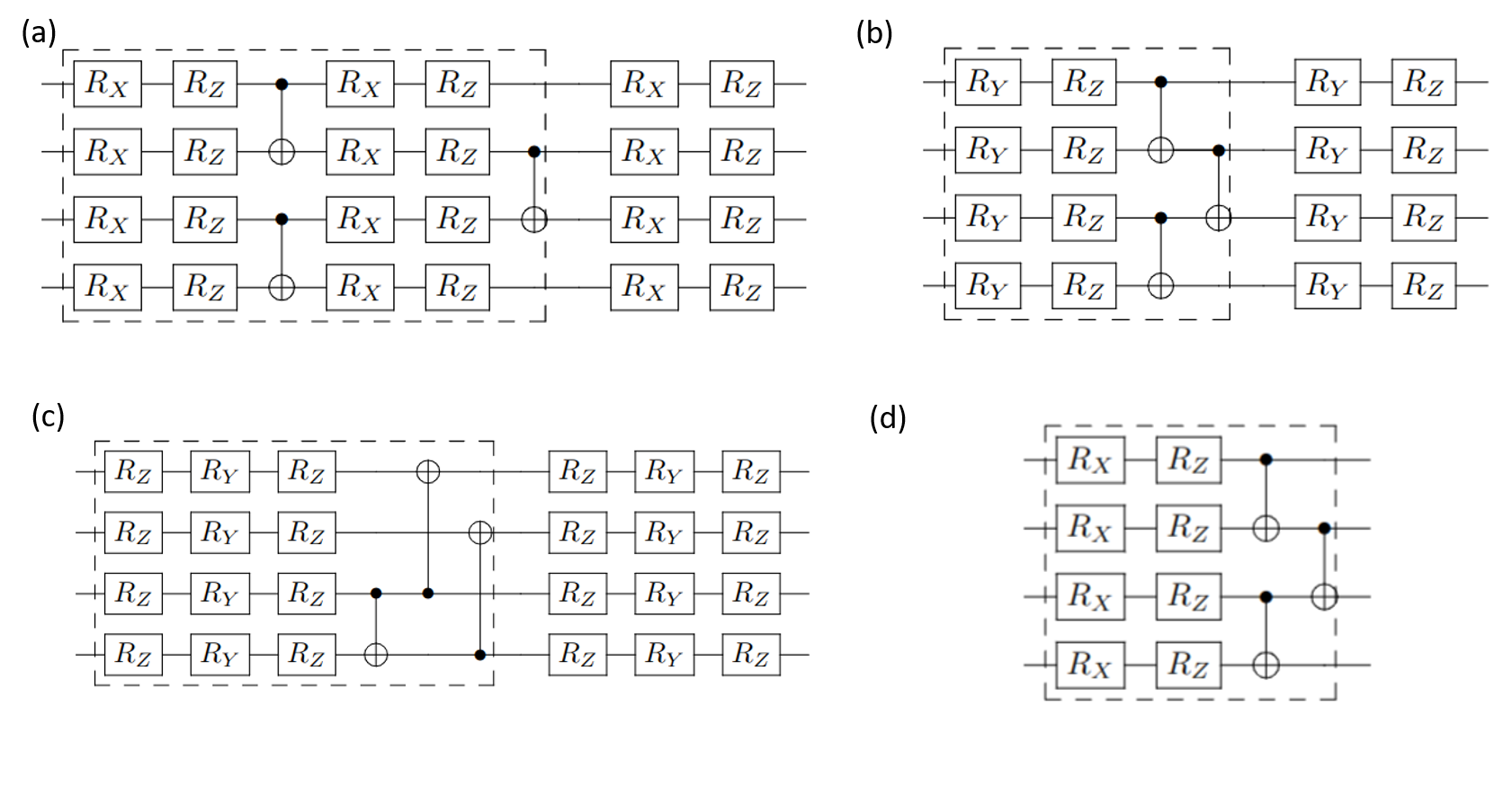}
    \caption{ Different ansatz templates used for numerical simulations. Ansatz templates 1 (a) and 2 (b) are taken from Ref. \cite{wu2021efficient}. We develop ansatz templates 3 (c) and 4 (d) to make some variation in the number of trainable gates in the template.
    The highlighted block in each template gets repeated for each layer.}
    \label{fig:ansatze}
\end{figure*}

\section{\label{sec:results}Results}
\subsection{\label{sec:lowerbound}Presence of a lower bound}
An analytical upper bound of the expressibility of an ansatz was presented in \cite{originalpaper} . It relates the expressibility of an ansatz to the complexity of the hypothesis space of the same. They use the covering number of the hypothesis space as a measure of the complexity. For $0<\epsilon\leq 1/10$, the covering number of the hypothesis space $\mathcal{H}$ is upper bounded by:
\begin{equation}
    \mathcal{N}(\mathcal{H},\epsilon,|\cdot|)\leq\left(\frac{7N_{gt}||\hat{O}||}{\epsilon}\right)^{d^{2k}N_gt}
\end{equation}
where $|\cdot|$ is the norm for the hypothesis space. It can be seen that the complexity of the hypothesis space (thus the expressibility of the ansatz used) depends on the type of gates (as directed by the parameter $k$) that we used to build said ansatz.
We followed a similar procedure as \cite{originalpaper} to get the lower bound to the covering number of the hypothesis space $\mathcal{H}$.
\begin{equation}\label{eq:bothbounds}
\left(\frac{3N_{gt}||\hat{O}||}{8\epsilon}\right)^{d^{2k}N_{gt}}\leq{N}(\mathcal{H},\epsilon,|\cdot|)
\end{equation}
The detailed derivation of the lower bound is provided in Appendix \ref{sec:Appendix}. Together with the upper bound, one can write the following inequality for the covering number of the hypothesis space of an ansatz, 
\begin{equation}
\label{eq:finalbothbound}
\left(\frac{3N_{gt}||\hat{O}||}{8\epsilon}\right)^{d^{2k}N_{gt}}\leq\mathcal{N}(\mathcal{H},\epsilon,|\cdot|)\leq\left(\frac{7N_{gt}||\hat{O}||}{\epsilon}\right)^{d^{2k}N_{gt}}
\end{equation}
From Equation~\ref{eq:finalbothbound},
We can characterize an unique property of the expressibility of an ansatz for a variational quantum algorithm by using the covering number of the Hypothesis space of an ansatz as a proxy \cite{originalpaper}. This equation implies the particular hypothesis space of an ansatz can only explore a limited region of the solution space, which is both upper and lower bounded, i.e., if the solution of a problem is lesser than the lower bound of the ansatz, the corresponding ansatz will never found the solution. This further enforces that construction of a VQA ansatz should be problem-specific.

\subsection{Numerical simulation}
The numerical simulations present in this paper have been done on Python on a system with 16 GB of RAM and a 2.40GHz processor with 4 Cores and 8 Logical Processors. We take help of the QISKIT library provided by IBM Quantum in order to construct the variational algorithm. 

We use the PySCF Driver in an `sto3g' basis to set up our Hamiltonian. Also, we use the Parity Mapper to encode into qubits. The NumPyMinimumEigensolver function is used to classically get the solution and the inbuilt VQE function of QISKIT simulated on a `statevector simulator' is used to run the variational code. The GradientDescent function of QISKIT is used as a classical optimizer with learning rate $0.4$ and tolerance $10^{-6}$ (in Hartree). We perform the simulations for the hydrogen molecule ($H_2$) and for four different ansatzes. The methodology that we follow for the ansatzes can be summarised in the following steps:
\begin{itemize}
  \item We take an ansatz and run our VQE algorithm for $H_2$ molecule at every atomic distance from $0.3$ to $2.1$ $\si{\angstrom}$ in steps of $0.2$. We obtain the minimum distance for each atomic distance. There is a distance at which the minimum energy is the lowest amongst all others. That would be the atomic distance according to the variational method. 
    \item We perform ten such trials. Now, it must be noted that this was done only for a single layer ( we call this the ansatz template \cite{entangling}). We increase the number of layers and repeat the same steps. We get an atomic distance and the atomic energy from each such endeavour.
    \item $H_2$ being a simple molecule, the corresponding Hamiltonian can be diagonalized classically. We get the actual atomic energy for $H_2$ at atomic distance by selecting the minimum eigenvalue of the Hamiltonian. 
    \item We use this value as reference and calculate the error for each layer used. The error is calculated for a single layer by taking the difference between the minimum VQE energy for each layer and the real energy. The average across ten trials gives us the error. We do this for each layer.
    \item Finally we plot an error vs layer number plot.
\end{itemize}
We do this for four different ansatz templates (Fig. \ref{fig:ansatze}).

\begin{figure}[hbt!]
    \centering
    \includegraphics[scale=0.4]{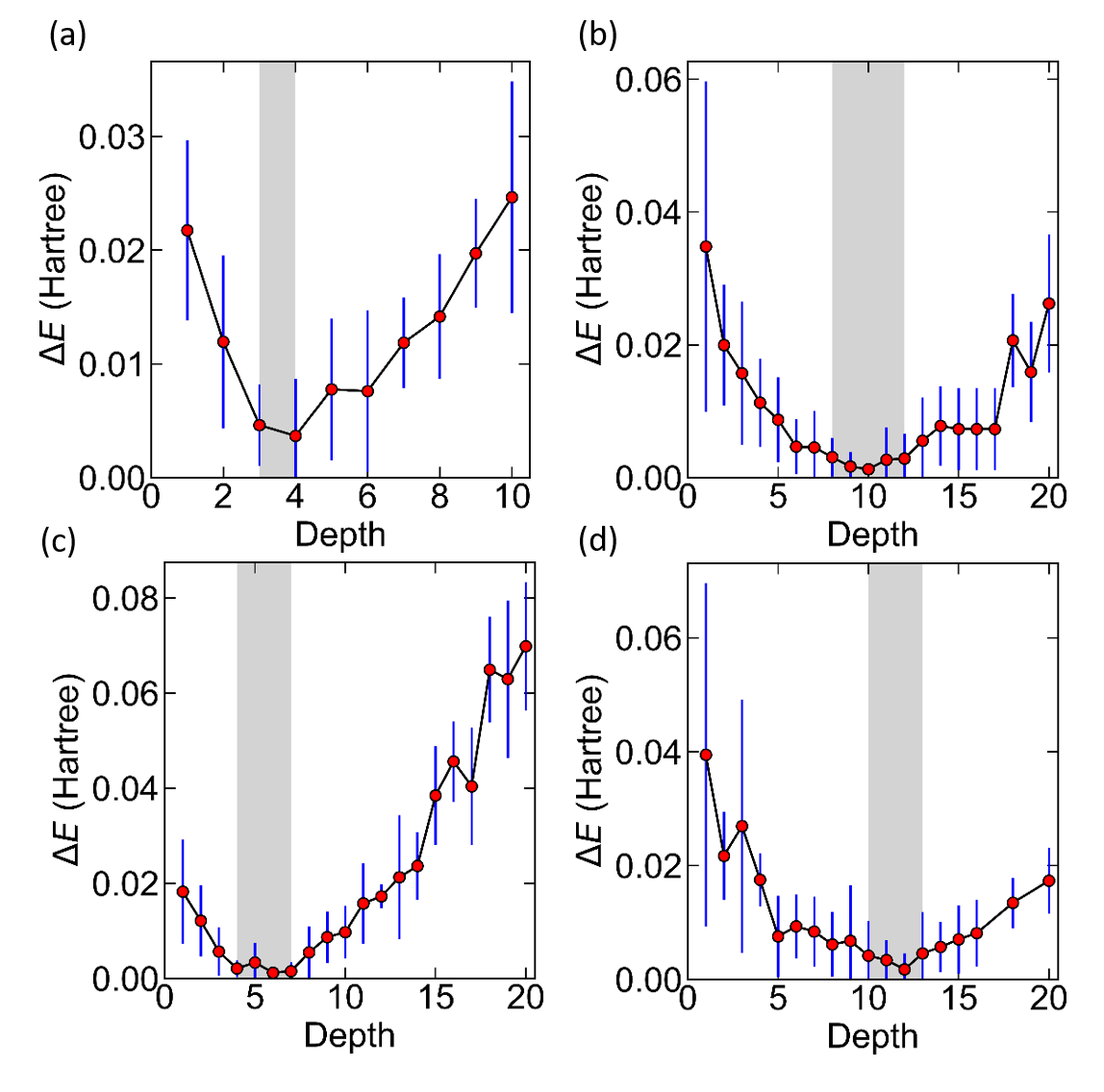}
    \caption{Variation of energy error with circuit depth (i.e., the number of repetitions of the circuit building block) for ansatz template (a) 1, (b) 2, (c) 3, and (d) 4 as depicted in Fig. \ref{fig:ansatze}. The error in energy ($\Delta E$)  is the deviation of the minimum energy obtained by our VQE algorithm from the minimum energy obtained by diagonalizing our $H_2$ Hamiltonian. The `Depth' is the number of times we have repeated our ansatz template. The blue vertical line at each datapoint indicates the error bar which give the standard deviation of each error point. For the point whose lower error bar takes it to negative values, we take a prior and set the lower bar to the mean itself. The curves dip for an intermediate depth and then rise up again. The plateau region, i.e.,the set of acceptable points in each ansatz template is highlighted in grey color.}
    \label{fig:simansatze}
\end{figure}
The results of our simulations are shown in Fig. \ref{fig:simansatze}. The first similarity that we notice across the four of them is that the error (i.e.,the difference between the VQE result and the actual result from diagonalization) for all of them seems to have a dip or a minimum point and then rises back up again with increase in depth of the ansatz template used. Each of the templates has a few layers for which it performs the best. Further increase or decrease of depth leads to a worse result. Ansatz template $1$ was repeated only till depth $10$ as it is the most complicated out of the four and is not expected to show a further dip. Furthermore, it is noted during the simulations that the time taken to execute the code for each ansatz template increases exponentially with increase in depth, as is expected.

\section{Discussions}
In order for an ansatz to perform well, there is a definite need to balance its expressibility and trainability \cite{zoeholmes}. For our case, the hypothesis space of the ansatz used for this particular problem ($\hat{O}$) is given by Equation \ref{eq:hypothesis}. As we have noted during our numerical simulations, our problem ($\hat{O}$) already has a solution (a minimum energy that we want to find), regardless of the ansatz used. How well our ansatz performs will be decided by how close it can get to the actual solution (the actual groundstate wavefunction in this case). This means that the hypothesis space of the ansatz must contain the solution for it to perform well.
\begin{figure*}[hbt!]
    \centering
    \includegraphics[scale=0.5]{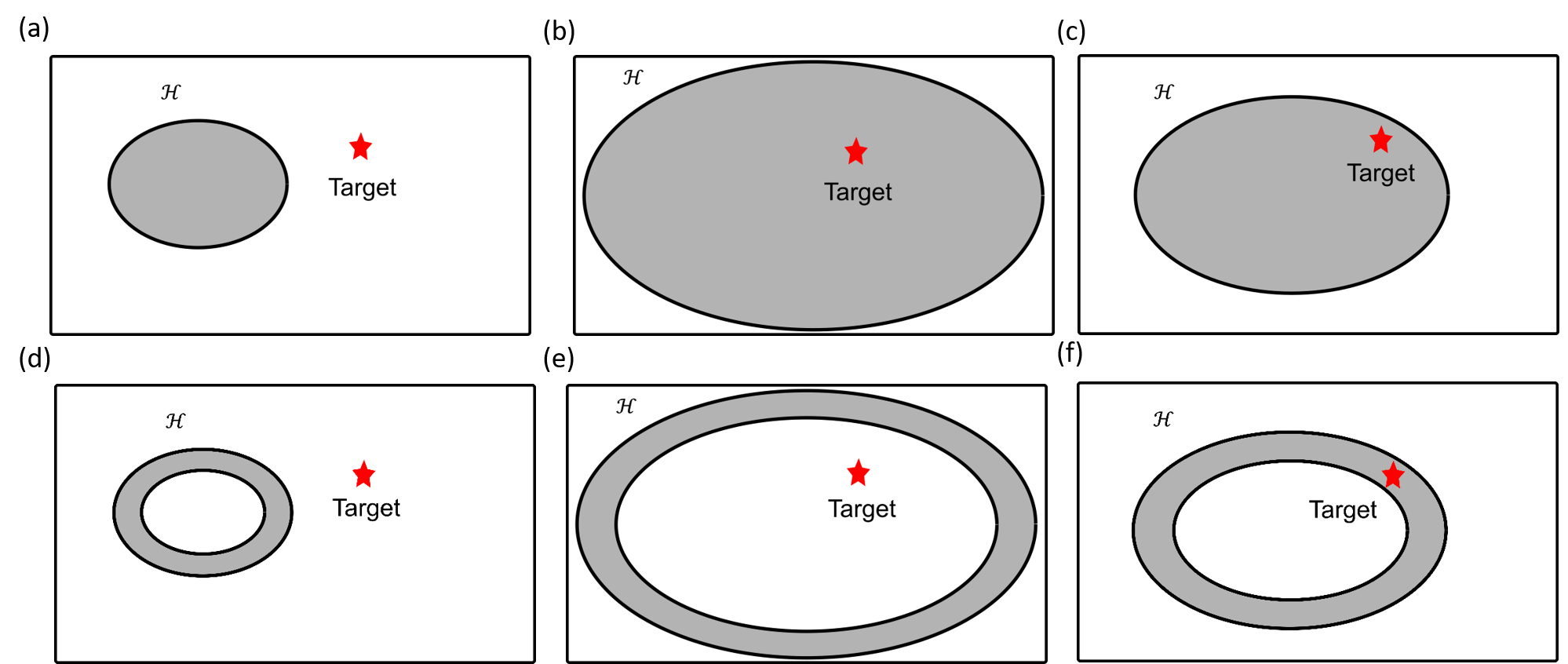}
    \caption{Schematic for the expressivity of an ansatz in hypothesis space $\mathcal{H}$ and its distance from the target solution in the same space. The expressivity of an ansatz is denoted by the shaded elliptic region, and the target is denoted by the red star. The upper panel describes (a) low, (b) high, and (c) moderate expressivity of an ansatz when the expressivity is defined by an upper bound only. The lower panel describes the (d) low, (e) high, and (c) moderate expressivity of an ansatz when a lower bound of expressivity also exists. (b) and (c) denotes that beyond a critical point the expressivity of an ansatz does not create any barrier in reaching the target. But (e) and (f) indicates that a very high expressive ansatz also never reach to the target because of the existence of its lower bound.}
    \label{fig:exp}
\end{figure*}

It was already known that the expressibility of an ansatz has an upper bound \cite{originalpaper}. The direct consequence of this was that the hypothesis space of the ansatz was thought to be a filled out object (like an ellipse) in some abstract space. When the expressibility of the ansatz used is low (Fig: \ref{fig:exp}(a)), the corresponding hypothesis space does not include the target and thus the ansatz would be unable to reach the correct solution. On the other hand, when the expressibility of the ansatz is very high (Fig. \ref{fig:exp}(b)), the hypothesis space is very big and the algorithm has to cover a lot of space in order to reach the target. The optimization algorithm becomes difficult to train along with increasing chances of encountering barren plateaus \cite{zoeholmes}. This again leads to a bad result. Thus, we can see that, in order to get a good solution (close to our actual target), we need an ansatz with moderate expressibility with a hypothesis space which covers the target but is not spread out too much from it (Fig. \ref{fig:exp}(c)) \cite{zoeholmes}.

But now, by including the lower bound too, we understand that the hypothesis space of the ansatz is not a continuous space. Instead, it can be thought of as an annular space. It has a lower bound below which the algorithm cannot search for the target solution. This is illustrated in Fig. \ref{fig:exp}(d)-\ref{fig:exp}(f). The presence of a lower bound to the expressibility of an ansatz tells us that if our solution corresponds to a wavefunction in a low expressive region then an initial choice of a highly expressive trial ansatz will never be able to reach the solution leading to an erroneous VQE result. The most important consequence of this endeavour is that as we can see from Fig. \ref{fig:exp}(e), the bad performance of an ansatz with a high expressibility is not just because of the difficulty of training but also due to its lack of expressibility. Hence, trainability and expressibility are intrinsically connected.

Since the logarithms of the bounds in Equation \ref{eq:finalbothbound} are easier to calculate numerically than the exponentials present, we prefer to work with the former.
\begin{equation}
\label{eq:finalboundlog}
\begin{split}
    &d^{2k}N_{gt}\log\left(\frac{3N_{gt}||\hat{O}||}{8\epsilon}\right) \\
    &\leq\log\mathcal{N}(\mathcal{H},\epsilon,|\cdot|) \\
    &\leq d^{2k}N_{gt}\log\left(\frac{7N_{gt}||\hat{O}||}{\epsilon}\right)
\end{split}
\end{equation}

\begin{figure}[!hbt]
    \centering
    \includegraphics[scale=0.4]{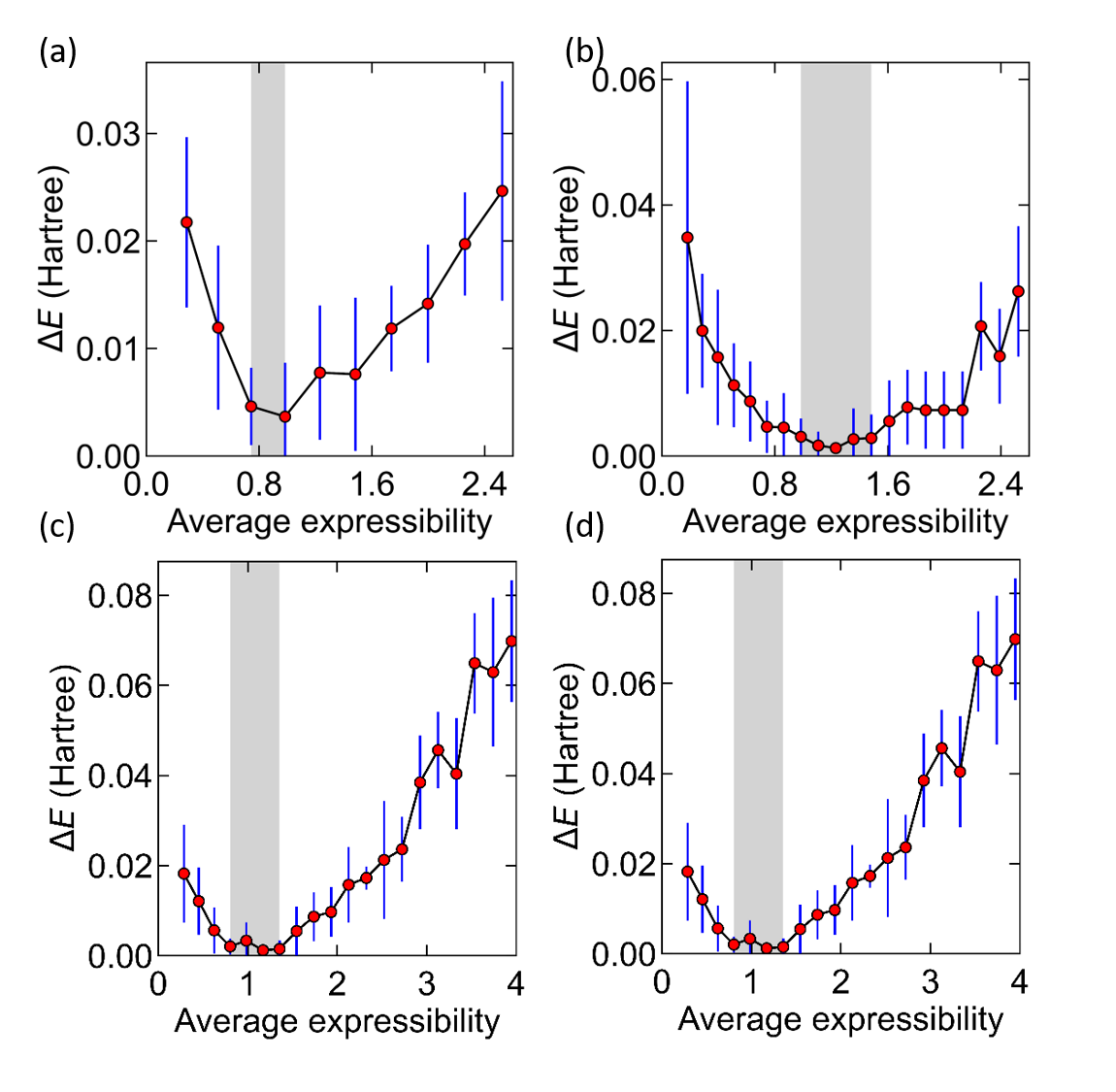}
    \caption{ Variation of error in energy ($\Delta E$) with average expressibility (in scales of $10^4$) for ansatz template (a) 1, (b) 2, (c) 3, and (d) 4. The x-axis from Fig:\ref{fig:simansatze} is replaced with our expression for average expressibility. The error seems to increase for low and high values of average expressibility. There are several points making up the minima of each curve. Ansatz depths with these values of average expressibility seem to perform the best (the least error). The set of acceptable points in each ansatz template is again highlighted in grey color.}
    \label{fig:avexpanstz}
\end{figure}

The only parameter in Equation \ref{eq:finalboundlog} that changes from ansatz to ansatz is the number of trainable gates $N_{gt}$. Plugging in the rest of the parameters, we get the bounds as a function of $N_{gt}$.
\begin{equation}
\label{eq:finalboundlognumbers}
\begin{split}
    &16N_{gt}\log\left(115.037956 N_{gt}\right)\\
    &\leq\log\mathcal{N}(\mathcal{H},\epsilon,|\cdot|)\\
    &\leq 16N_{gt}\log\left(43.8239831 N_{gt}\right)
\end{split}
\end{equation}
The number of trainable gates changes with the layer number differently for different ansatz templates, as we tabulate in Table \ref{tab:ngtdiff}.
\begin{table}[htbp]
\caption{Ansatz templates and trainable gates in them.}
\begin{tabular}{p{2cm}p{3cm}p{2cm}}
\hline\hline
Ansatz & Single Layer $N_{gt}$ & $N_{gt}$ for layer $n$\\ \hline
1             & $3 \times 4 + 3\times 4=12$              & $12n +12$               \\ \hline
2             & $2 \times 4 + 2 \times 4 = 16$     & $8n+8$\\ \hline
3             & $4 \times 4 + 4 \times 2 = 24$     & $16n + 8$  \\ \hline
4             & $2 \times 4 = 8$          & $8n$             \\ \hline\hline
\end{tabular}
\label{tab:ngtdiff}
\end{table}
 We can use this to get a numerical value for the upper bound and the lower bound of the expressibility of each layer for every ansatz. The first thing that it allows us to do is to define an average expressibility value for each layer of the ansatz. This completes the search for a general expression for expressibility of an ansatz and  is also a much more easily obtainable value than what was put forward in previous literature \cite{entangling,alternatingexp}. We then change the x-axes in Fig. \ref{fig:simansatze} in order to get an error vs average expressibility graph for all four of our ansatz templates. As we can see in Fig. \ref{fig:avexpanstz}, the error of our result obtained from the variational method is more for high and low average expressibility, but the error is less for an intermediate values of average ansatz expressibility. This agrees with previous results \cite{zoeholmes}.

 We notice that for each ansatz template, there are certain layers for which the error is the least among all others. We call these set of points the set of acceptable points. Next, we define the expressibility range from the upper bound of the rightmost acceptable point and the lower bound of the leftmost acceptable point in Fig: \ref{fig:simansatze} as the best expressive range for each of the ansatz templates. For these set of layers, the ansatz template performs the best over the entire range of repetitions. Even if we keep on adding more layers, we will not get a better performance. Secondly, we also calculate the average error over these sets of acceptable points for the four ansatz templates that we have. The result is that, as we have noted in Table \ref{tab:avgerrexpreg}, we notice a direct connection between the span of the best expressive range and the average error. For low average error across the set of acceptable points, the span of the best expressive range is the widest. There is a direct correlation between the two.
\begin{table}[htbp]
\caption{We tabulate the span of the best expressive range (defined as the difference between the average expressibility of the rightmost acceptable point and the lower bound of the leftmost acceptable point) and the average error over this range for the four different ansatz templates. The best expressive range is selected by choosing the points with the lowest error points from Fig. \ref{fig:avexpanstz}. We can see a direct correlation between the span of the best expressive range and the average error over the range. They are inversely proportional. The ansatz template that performs the best does so over a wider range of layers than the ones that perform poorly.}
\begin{tabular}{p{2cm}p{3cm}p{3cm}}
\hline
\hline
Ansatz & Range & Average error \\ \hline
$1$             & $3399.50$             &  $0.004145$               \\ \hline
$2$             & $6342.89$     & $0.002861$\\ \hline
$3$             & $6740.02$    & $0.002046$ \\ \hline
$4$             & $5176.76$          & $0.003473$             \\ \hline\hline
\end{tabular}
\label{tab:avgerrexpreg}
\end{table}


\section{\label{sec:conclusions}Conclusions}
In this work, we have shown that there exists a lower bound to the covering number of the hypothesis space of an ansatz. We have found a correlation between the aforementioned lower bound to the complexity of the hypothesis space and correspondingly to the expressibility of the ansatz used. This changes how the hypothesis space exists in an abstract space. By our findings, it is more close to an annular region than a continuous space. This makes the expressibility and trainability of an ansatz more inter-wined with the ansatz performances \cite{zoeholmes,entangling}.
We have also performed numerical simulations to plot  the error of a variational algorithm vs the layer number of the ansatz template. We noticed that the intermediate layer numbers which have the moderate expressibility, the error is the lowest,  We have also found that the average error across the set of acceptable points and the best expressive region for an ansatz template are inversely proportional.

Our analytic result and numerical simulations have been done in a noise-free situation. The next step should be to extend it to include noise. Further, in this paper we have only considered expressibility as a measure of ansatz performance to link our results with  trainability. However, it would be interesting to look at expressibility and trainability together as a single evaluation parameter in order to better judge ansatzes.Our results show that the performance of an ansatz is an intrinsic property of the "ansatz and problem" system and not just an issue of trainability. We hope that, in the future, this result can be used to select better ansatzes for practical application of Variational Quantum Algorithms.
We sincerely thank Mandar Thatte, ICT Bombay for insightful discussions.


\bibliographystyle{unsrt}

\appendix
\section{Derivation of the lower bound}\label{sec:Appendix}
We follow a similar process as in \cite{originalpaper} to find a lower bound to the covering number of the hypothesis space.Let $S$ be the $\epsilon$-covering set for the set $\mathcal{U}(d^k)$.
\begin{equation}
\label{eq:coverings}
    |S|=\mathcal{N}(\mathcal{U}(d^k),\epsilon,||\cdot||)
\end{equation}
We want to find the covering set for the set $\mathcal{H}_{circ}$. For this, we define another set $\Tilde{S}$ such that:
\begin{equation}
    \Tilde{S}:=\{\prod_{i\in\{N_{gt}\}}\hat{u}_i(\theta_i)\prod_{j\in(N_g -N_{gt})}\hat{u}_j|\hat{u}_i(\theta_i)\in S\}
\end{equation}
The ordered $\hat{u}_i$ in $\Tilde{S}$ are  the particular ones used in the operator group $\mathcal{H}_{circ}$. Now, we show that  $\Tilde{S}$ is an `$\epsilon '$-covering set' of $\mathcal{H}_{circ}$ and find $\epsilon '$. First, we replace the $N_{gt}$ trainable gates $\hat{u}_i(\theta_i)$ in $\Tilde{S}$ with the nearest element in $S$ (all $\epsilon$ distance away as $S$ is $\epsilon$-covering set of $\mathcal{U}(d^k)$) . Now, we have :
\begin{equation*}
    \hat{U}_{\epsilon}(\theta)\in\Tilde{S}
\end{equation*}
We will now prove that for any element $a\in\mathcal{H}_{circ}$ and element $b\in \Tilde{S}$, $d(a,b)\leq \epsilon'$, where $d$ is the distance measure for the operator group i.e.,$||\cdot||$. It can be shown that the distance measure satisfies:
\begin{equation}
\label{eq:distancehcirctildes}
\begin{split}
    &||\hat{U}(\theta)^\dag \hat{O} \hat{U}(\theta)-\hat{U}_{\epsilon}(\theta)^\dag \hat{O} \hat{U}_{\epsilon}(\theta)||\\
    & \leq||\hat{U}-\hat{U}_\epsilon||\;||\hat{O}||\\
    & \leq N_{gt}||\hat{O}||\epsilon=\epsilon'
\end{split}
\end{equation}
where we use the triangle inequality for the first inequality and $||\hat{U}-\hat{U}_\epsilon ||\leq N_{gt}\epsilon$ for the second inequality. From our earlier definition, we can say that the set $\Tilde{S}$ is an `$N_{gt}||\hat{O}||\epsilon$ -covering set' for $\mathcal{H}_{circ}$. Thus:
\begin{equation}
\label{eq:sbarn}
    |\Tilde{S}|=\mathcal{N}(\mathcal{H}_{circ},N_{gt}||\hat{O}||\epsilon,||\cdot||)
\end{equation}
From Equations \ref{eq:Uupperlower} and \ref{eq:coverings}, we have:
\begin{equation}
\label{eq:slower}
    \left(\frac{3}{4\epsilon}\right)^{d^{2k}}\leq|S|
\end{equation}
We must remember that we replaced $N_{gt}$ trainable gates in $\Tilde{S}$ using gates from $S$ and we have $|S|^{N_{gt}}$ combinations for the gates in $\Tilde{S}$. This gives us:
\begin{equation}
\label{eq:lowerfors}
    \left(\frac{3}{4\epsilon}\right)^{d^{2k}N_{gt}}\leq|\Tilde{S}|
\end{equation}
Finally, we use Equations \ref{eq:sbarn} and \ref{eq:lowerfors} to get a lower bound to the covering number of the operator group $\mathcal{H}_{circ}$:
\begin{equation}
    \left(\frac{3}{4\epsilon}\right)^{d^{2k}N_{gt}}\leq\mathcal{N}(\mathcal{H}_{circ},N_{gt}||\hat{O}||\epsilon,||\cdot||)
\end{equation}
We rescale $\epsilon$ by using $\epsilon=\frac{2\epsilon}{N_{gt}||\hat{O}||}$
\begin{equation}
\label{eq:lowerb}
    \left(\frac{3N_{gt}||\hat{O}||}{8\epsilon}\right)^{d^{2k}N_{gt}}\leq\mathcal{N}(\mathcal{H}_{circ},2\epsilon,||\cdot||)
\end{equation}
Next, we draw a relation between the covering number of the operator space $\mathcal{H}_{circ}$ and the hypothesis space $\mathcal{H}$. In order to do that, it can be shown from \cite{originalpaper} that a Bi-Lipschitz mapping exists between the two spaces.

Let ($\mathcal{H}_1,d_1$) and ($\mathcal{H}_2,.d_2$) be two metric spaces and the map $f$: $\mathcal{H}_1 \rightarrow \mathcal{H}_2$ be Bi- Lipschitz with:
\begin{equation}
\label{eq:lemma}
    \begin{split}
        d_2(f(x),f(y))&\leq Kd_1(x,y); \forall x,y\in \mathcal{H}_1\\
         d_2(f(x),f(y))&\geq kd_1(x,y); \forall x,y\in \mathcal{H}_1 \;with \;d_1(x,y)<r
    \end{split}
\end{equation}
\par Then their respective covering numbers follow the relations:
\begin{equation}
\label{eq:coveringnumberrealtion}
    \mathcal{N}(\mathcal{H}_1,\frac{2\epsilon}{k},d_1)\leq\mathcal{N}(\mathcal{H}_2,\epsilon,d_2)\leq\mathcal{N}(\mathcal{H}_1,\frac{\epsilon}{K},d_1)
\end{equation}
where the left inequality requires 
\begin{equation}
    \epsilon\leq\frac{k r}{2}
\end{equation}

In our case,
\begin{equation}
    \mathcal{H}_1=\mathcal{H}_{circ}
\end{equation}
where $\mathcal{H}_{circ}=\{\hat{U}(\theta)^\dag\hat{O} \hat{U}(\theta)|\theta\in \Theta \}$ is the operator group with the corresponding distance metric $d_1=||\cdot||$.
\par The other space:
\begin{equation}
    \mathcal{H}_2=\mathcal{H}
\end{equation}
where $\mathcal{H}=\left\{Tr(\hat{U}(\theta)^\dag \hat{O}\hat{U}(\theta)\rho)|\theta\in \Theta\right\}$ is the hypothesis space with the distance metric $d_2=|\cdot|$.

In order to find $K$, we follow the similar process of replacing the gates used in the ansatz $\hat{U}(\theta)$ with the nearest gates in the covering set $S$ to get a new ansatz $\hat{U}_\epsilon(\theta)$. This is used in the first inequality in Equation \ref{eq:lemma} to give \cite{originalpaper}: 
\begin{equation}
\begin{split}
    d_2(Tr(\hat{U}_\epsilon(\theta)^\dag \hat{O}\hat{U}_\epsilon(\theta)\rho)-Tr(\hat{U}(\theta)^\dag \hat{O}\hat{U}(\theta)\rho))&=\\d_1(\hat{U}_\epsilon(\theta)^\dag \hat{O}\hat{U}_\epsilon(\theta)-\hat{U}(\theta)^\dag \hat{O}\hat{U}(\theta))&
    \end{split}
\end{equation}
hence
\begin{equation}
    K=1
\end{equation}

For a Bi-Lipshitz map, it can be shown \cite{referthesis} that through isometry of the map:
\begin{equation}
    k=\frac{1}{K}
\end{equation}
Thus, since $K=1$, hence it follows that $k=1$. It is possible for the Bi-Lipshitz to have a value $k > \frac{1}{K}$, which will further tighten \ref{eq:lowerb}, however, in this work we have focused primarily on finding a lower bound of the expressibility of an ansatz.
Using this and the left inequality in Equation~\ref{eq:coveringnumberrealtion}, we get:
\begin{equation}
    \mathcal{N}(\mathcal{H}_{circ},2\epsilon,||\cdot||)\leq\mathcal{N}(\mathcal{H},\epsilon,|\cdot|)
\end{equation}
Finally, using Equation \ref{eq:lowerb}, we get the lower bound to the covering number of the hypothesis space $\mathcal{H}$.
\begin{equation}
\left(\frac{3N_{gt}||\hat{O}||}{8\epsilon}\right)^{d^{2k}N_{gt}}\leq{N}(\mathcal{H},\epsilon,|\cdot|)
\end{equation}

It must be noted that the left inequality in Equation \ref{eq:coveringnumberrealtion} requires $\epsilon\leq\frac{k r}{2}$ where $d_1(x,y)\leq r$. Hence, we can use the result from the Equation \ref{eq:distancehcirctildes} to get:
\begin{equation}
    r=N_{gt}||\hat{O}||\epsilon
\end{equation}
Plugging all of this into the condition, we get:
\begin{equation}
\label{eq:lowerboundonngt}
    N_{gt}\geq\frac{2}{||\hat{O}||}
\end{equation}
This gives us a lower bound on the number of trainable gates that we need in our ansatz for our result to be valid. For our case, $d=2$ (as we have qubits) , $k=2$ (as only single and two qubit gates are used in our ansatz templates). We know that the parameter $\epsilon$ can take values between $0$ and $\frac{1}{10}$. We select $\epsilon=\frac{1}{100}$. The operator norm for the Hamiltonian is different for different atomic distances. We work with its value when $d=0.8$(Angstrom) as the energy is always minimum between $d=0.7$ and $d=0.9$ for $H_2$. Thus we have $||\hat{O}||=||H||=1.16863955$. Additionally, plugging $||H||$ into Equation \ref{eq:lowerboundonngt}, we get the trivial lower bound on the number of trainable gates we need as $N_{gt}\geq 1$.  

\end{document}